\documentclass[aps,prl,twocolumn,showpacs,superscriptaddress,reprint]{revtex4-1}
\usepackage{amsmath}
\usepackage{amsfonts}
\usepackage{amssymb}
\usepackage[dvipdfmx]{graphicx}
\usepackage[usenames,dvipsnames]{color} 
\usepackage[normalem]{ulem}

\newcommand{\kb}{k_{\mbox{\scriptsize B}}}
\newcommand{\lgle}{\left\langle}
\newcommand{\rgle}{\right\rangle}
\newcommand{\Fs}{\Phi_S}
\newcommand{\Fl}{\Phi_L}
\newcommand{\Fselfs}{\Phi_{s, S}}
\newcommand{\Fselfl}{\Phi_{s, L}}

\begin{document}

\title{Cluster Glass Transition of Ultrasoft-Potential Fluids at High Density}

\author{Ryoji Miyazaki}
\author{Takeshi Kawasaki}
\author{Kunimasa Miyazaki}
\email[Corresponding author. ]{\\ miyazaki@r.phys.nagoya-u.ac.jp}
\affiliation{Department of Physics, Nagoya University, Nagoya 464-8602, Japan}

\date{\today}

\begin{abstract}
Using molecular dynamics simulation, we investigate the slow dynamics of a supercooled binary mixture of 
soft particles interacting with a generalized Hertzian potential.
At low density, it displays typical slow dynamics near its glass
 transition temperature. At higher densities, particles bond together, forming clusters, and the
 clusters undergo the glass transition.  
 The number of particles in a cluster increases one by one 
as the density increases.
We demonstrate that there exist the multiple cluster-glass phases characterized by a different number of particles per cluster, 
each of which is separated by distinct minima. 
Surprisingly, a so-called higher order singularity of the mode-coupling theory
 signaled by a logarithmic relaxation is observed in the
 vicinity of the boundaries between monomer and cluster glass phases.
The system also exhibits rich and anomalous dynamics in the cluster glass phases, 
such as the decoupling of the self- and collective dynamics.
\end{abstract}


\maketitle

Ultrasoft potential systems have attracted special attention in the soft matter community
in the last two decades~\cite{Malescio2007,Likos2006b}.
The ultrasoft potentials are pairwise isotropic and repulsive interactions whose value remains finite even if
the particles fully overlap each other.
Typical examples include the Gaussian, harmonic, and Hertzian
potentials. 
Systems with such potentials exhibit counterintuitive thermodynamic behaviors depending
on the shape of the Fourier spectrum $\tilde{v}(k)$ of the potential $v(r)$.
According to a criteria proposed by Likos {\it et al.}~\cite{Likos2001b}, 
if $\tilde{v}(k)$ is positive definite, which is referred to as the $Q^{+}$ class, the system undergoes a reentrant
transition, {\it i.e.}, the fluid-crystalline phase boundary in the temperature-density plane has a maximal peak,
so that the once crystallized solid melts again as the density increases~\cite{Stillinger1976,Pamies2009}.
On the other hand, if $\tilde{v}(k)$ is negative at finite wave vectors, which is called the $Q^{\pm}$ class~\cite{Likos2001b}, 
the reentrant melting at high density disappears and instead the particles bond together and form a cluster crystal,
where each lattice site is occupied by several overlapped
particles.
Since the first sign of such coagulation has been reported in a numerical simulation~\cite{Klein1994},
the existence of cluster crystal phases has been demonstrated unambiguously for model
ultrasoft potential
systems~\cite{Likos2001b,Mladek2006,*Likos2007,Moreno2007b,Lascaris2010,Zhang2010c,Coslovich2011sm,*Montes-Saralegui2013jpcm,Miller2011sm}. 
Most of the studies focused on  
the generalized exponential model (GEM), defined by 
$v(r) =\epsilon \exp\left[- (r/\sigma)^{n}\right]$ with
$n>2$, where $\epsilon$ and $\sigma$ set the energy and length scales
~\cite{Moreno2007b,Zhang2010c,Montes-Saralegui2013jpcm,Mladek2006,*Likos2007}.
Similar cluster crystals were also found for more realistic
systems~\cite{Lenz2012prl,Sciortino2013nat,Bernabei2013sm,*Slimani2014acsml} and even for the 
quantum supersolids~\cite{Cinti2014natc,Diaz-Mendez2015pre}.  

The glass transition of ultrasoft potential fluids was also studied~\cite{Berthier2010i,Ikeda2011,*Ikeda2011jcp2,Coslovich2012jcp}.
The reentrant behavior of glass phase diagrams has been reported for 
the $Q^{+}$ class fluids such as the Gaussian~\cite{Ikeda2011,Ikeda2011jcp2}, harmonic,
and the Hertzian potential systems~\cite{Berthier2010i,Zhao2011prl,*Wang2012sm,PicaCiamarra2013sm}.
On the other hand, for the $Q^{\pm}$ class systems, the cluster glasses were found in a binary mixture of the GEM fluid at
high density~\cite{Coslovich2012jcp}.   

In this Letter, we numerically study ultrasoft particles interacting with the generalized Hertzian
potential (GHP) defined by
\begin{equation}
 v(r) = \frac{\epsilon}{\alpha} \left( 1 - \frac{r}{\sigma} \right)^{\alpha} \hspace*{1.0cm}\mbox{for } r \leq \sigma
\label{eq:gHertz}
\end{equation}
and $v(r)=0$ for $r > \sigma$.
This model has been studied extensively for $\alpha=2$ (harmonic) and $\alpha=5/2$
(Hertzian) in the context of both glass and jamming
transitions~\cite{Berthier2010i,Wang2012sm,PicaCiamarra2013sm}.
They belong to the $Q^{+}$ class, because $\tilde{v}(k) > 0$.
If $\alpha< 2$, $\tilde{v}(k)$ becomes negative at finite wave vectors and thus it falls in the 
$Q^{\pm}$ class~\cite{SM}. 
Recently, thermodynamic properties of a two dimensional monatomic GHP model have been studied
and it was found that, instead of the clusters of multiply overlapping particles on the lattice sites,
the system displays varieties of crystalline structures, such as anisotropic lanes, 
honeycomb lattices, and even the kagome lattices, followed by a cascade of more complex
superlattice phases at higher densities~\cite{Miller2011sm}.
It is surprising that a deceptively simple model exhibits such exotic ground states, 
although their existence is not theoretically excluded in the $Q^{\pm}$ class fluids.
The next natural question is whether the dynamics, especially the glassy slow dynamics, of the GHP fluid is
as anomalous as its thermodynamic properties. 

\begin{figure}[t]
 \begin{center}
\includegraphics[width=0.5\textwidth,angle=0]{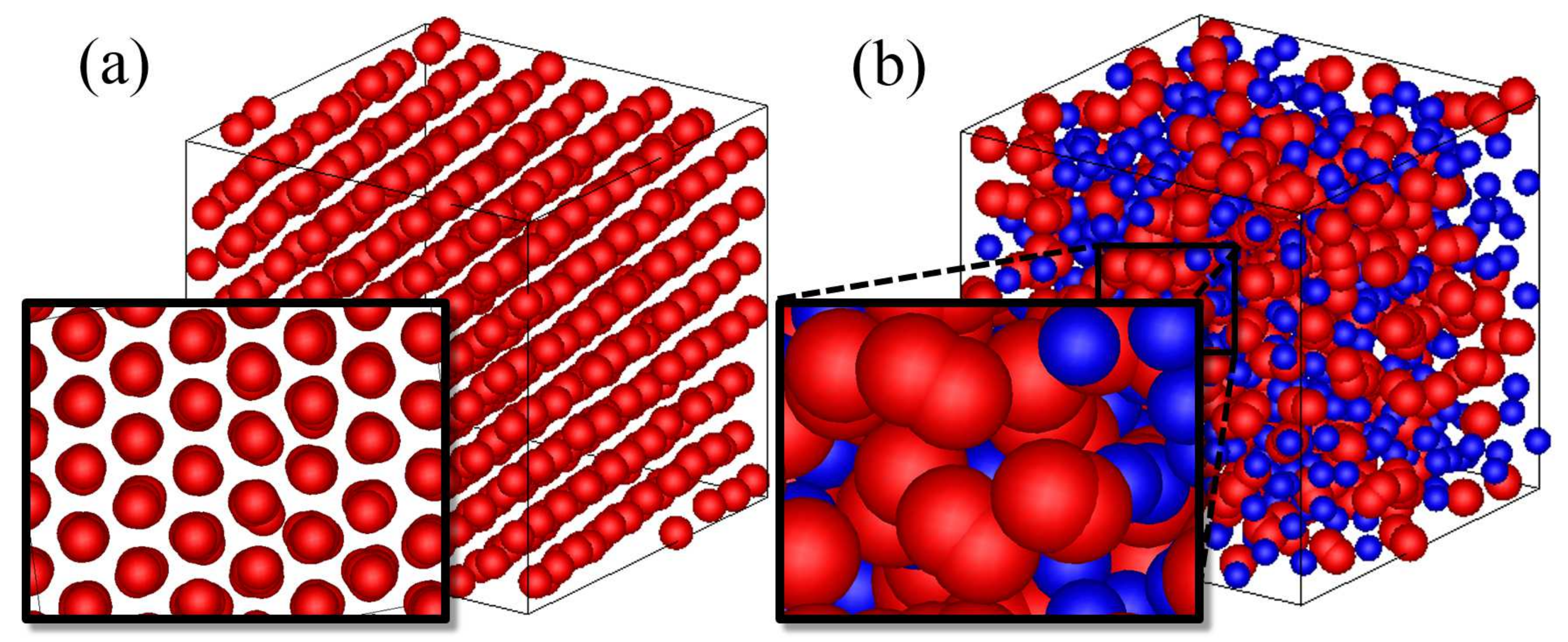}
 \end{center}
 \caption{Snapshots of the GHP system with $\alpha=3/2$.
The particle sizes are not drawn to scale but changed to optimize the visibility.
 (a) A hexagonal phase for the monatomic system at $\rho = 2.6$ and $T = 50$.
 Particles are seen to align along the columns.
 The inset is a top view seen from the direction parallel to the columns. 
 (b) A cluster glass phase of the binary system with large (red) and small (blue) particles at $\rho = 1.3$ 
and $T = 98$.
 The inset is the close-up.
}
 \label{Fig1}
\end{figure}
For this goal, we perform the molecular dynamics (MD) simulation for the 50:50 binary mixture of 
large ($L$) and small ($S$) particles with the same mass, $m$, interacting with the GHP in three dimensions.
In Eq.~(\ref{eq:gHertz}), we define the length scale $\sigma$ as 
$\sigma_{ab}= (\sigma_{a} + \sigma_{b})/2$, where
$\sigma_a$ is the particle diameter of the $a$ ($=L$, $S$) component.
We set $\sigma_S$, $m$, $10^{-4}\epsilon/\kb$ ($\kb$ is the Boltzmann constant), 
$\sqrt{m \sigma_S^{2}/\epsilon}$, as the units of
the length, mass, temperature, and time, respectively. 
We mainly study the system of $N=1000$ particles. 
We checked that the system size effect is small.
Simulations were performed in the $NVE$ ensemble with the velocity Verlet algorithm 
with periodic boundary conditions. 
We use the temperature $T$ and the number density $\rho$
as the control parameters. 
We run simulations for $\alpha=3/2$, $1.7$, $2$, and $5/2$ for a wide range of density between 
$0.5 \leq \rho \leq 3$.
We present mainly the results for $\alpha=3/2$.
We first run several MD simulations of the monatomic system, {\it i.e.},  $\sigma_{S} =\sigma_{L}$, 
with $\alpha=3/2$ to check whether there exists any exotic crystalline phase which may underlie an exotic glass
transition of the binary counterpart.
Only the non-fcc/bcc crystalline phase which we found up to the largest density ($\rho = 3$) was  
the hexagonal crystal, {\it i.e.}, stringlike columns aligned in the hexagonal lattice, as 
shown in Fig.~\ref{Fig1}(a).
This indicates that there exists no cluster crystal phases in this density window, 
although more elaborate analysis using thermodynamic integration would be necessary to determine
the accurate phase diagram~\cite{Pamies2009}.

\begin{figure}[tb]
 \begin{center}
\includegraphics[width=0.5\textwidth,angle=0]{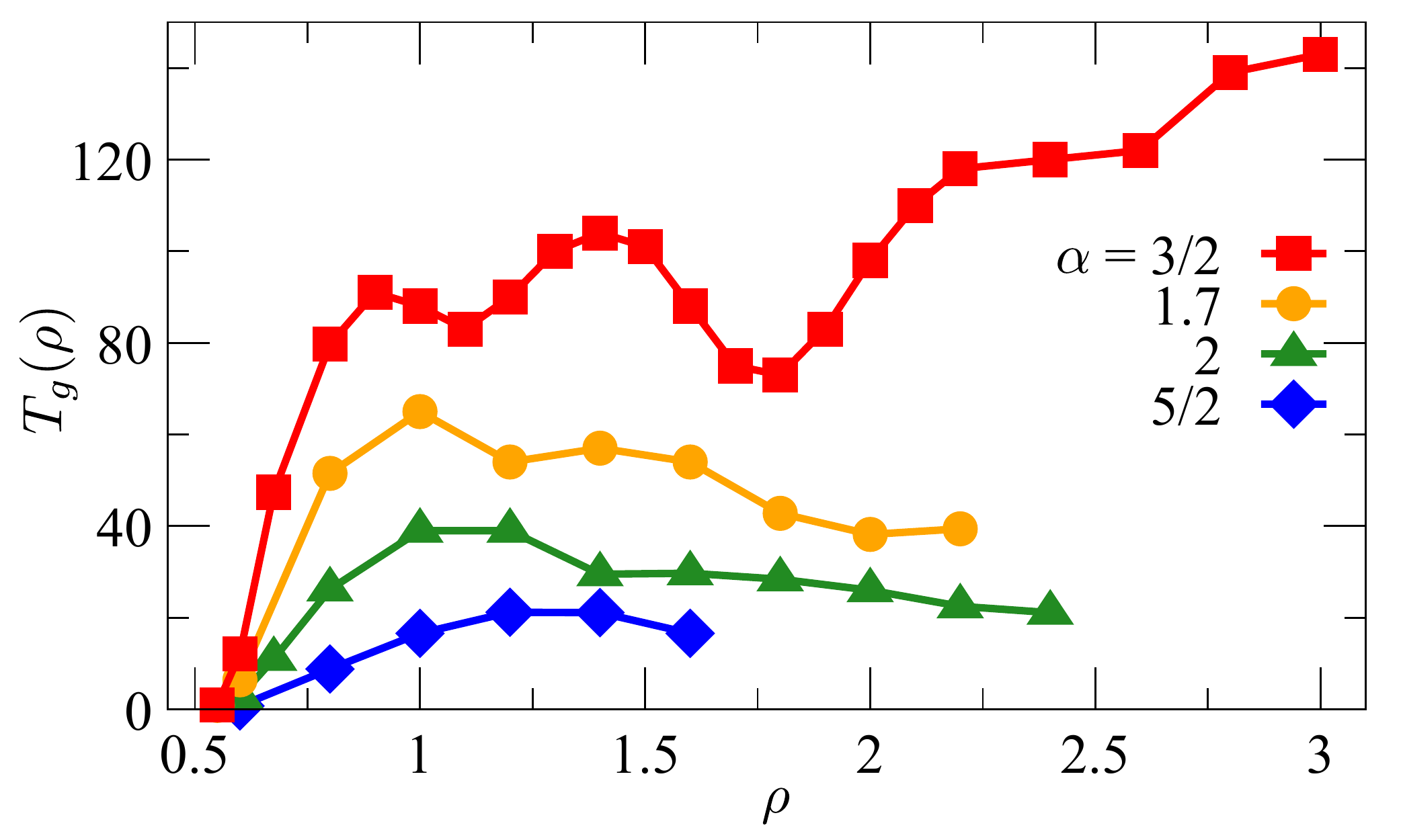}
 \end{center}
\vspace*{-0.5cm}
 \caption{The glass  lines $T_g(\rho)$ (or iso-$\tau$ lines) 
for 
$\alpha=3/2$ (square), $1.7$ (circle), $2$ (triangle), and $5/2$ (diamond).
}
 \label{Fig2}
\end{figure}

We now turn to the binary mixture and analyze its structural and dynamical  properties.  
The size ratio of the particles, $\sigma_{L}/\sigma_{S}$, is set to 1.4 in order to
avoid crystallization.
We determine the glass phase diagram from the relaxation time $\tau$ of 
the collective density correlation functions 
defined by 
$\Phi_{a}(k,t)=\lgle \rho_{a}(k,t)\rho_{a}(-k,0)\rgle/\lgle |\rho_{a}(k,0)|^2\rgle$,  
where
$\rho_{a}(k,t)$ is the collective density fluctuation of the
$a$ ($=\mbox{L}, \mbox{S}$) component.
We define $\tau$ by $\Phi_{a}(k, \tau) =0.1$ at the wave vectors
$k\approx 2\pi/\sigma_{a}$ corresponding to the nearest neighbor peak 
of the corresponding static structure factor.  
For a fixed density between $0.5 \lesssim \rho \lesssim 3$,
we search for the temperature at which $\Phi_a(k,t)$ develops glassy slow dynamics. 
We define the glass transition point $T_g$ as the temperature at which $\tau$ of
$\Fl(k,t)$ reaches $10^3$.  
Figure~\ref{Fig2} shows the iso-$\tau$ line, or the glass transition line $T_g(\rho)$, for several $\alpha$'s.
$T_g$ determined by standard methods, such as the Adam-Gibbs extrapolation, lies parallel but slightly below this line.  
For $\alpha \geq 1.7$, the systems crystallized at high densities, so that the glass lines
terminate there as shown in Fig.~\ref{Fig2}.
For $\alpha=5/2$, $T_g(\rho)$ shows reentrant behavior and almost matches with the result
for a polydisperse system~\cite{Berthier2010i}, implying
that the result is insensitive to the size dispersity of particles.
The overall shape of $T_g(\rho)$, including the position of the reentrant peak, is quite similar to the
fluid-crystal binodal line for the monatomic system, although the latter lies at higher temperatures~\cite{Pamies2009}. 
As $\alpha$ decreases, $T_g(\rho)$ tends to increase and the reentrance behavior is suppressed. 
For $\alpha = 3/2$, the reentrant behavior at high density disappears and $T_g(\rho)$ develops several cusplike
minima observed at $\rho\approx 1.1$, $1.8$, and $2.5$.
The observed minima are a clear sign that the system forms a new class of glasses.
Inspecting the system near the glass line by eye, we find that
all particles remain monomers below the first minimum $\rho \lesssim 1.1$,
but for $\rho \gtrsim 1.1$ the large particles bond together forming dimer-shaped
clusters, as shown in the snapshot of Fig.~\ref{Fig1}(b). 

\begin{figure}[b]
 \begin{center}
\includegraphics[width=\columnwidth]{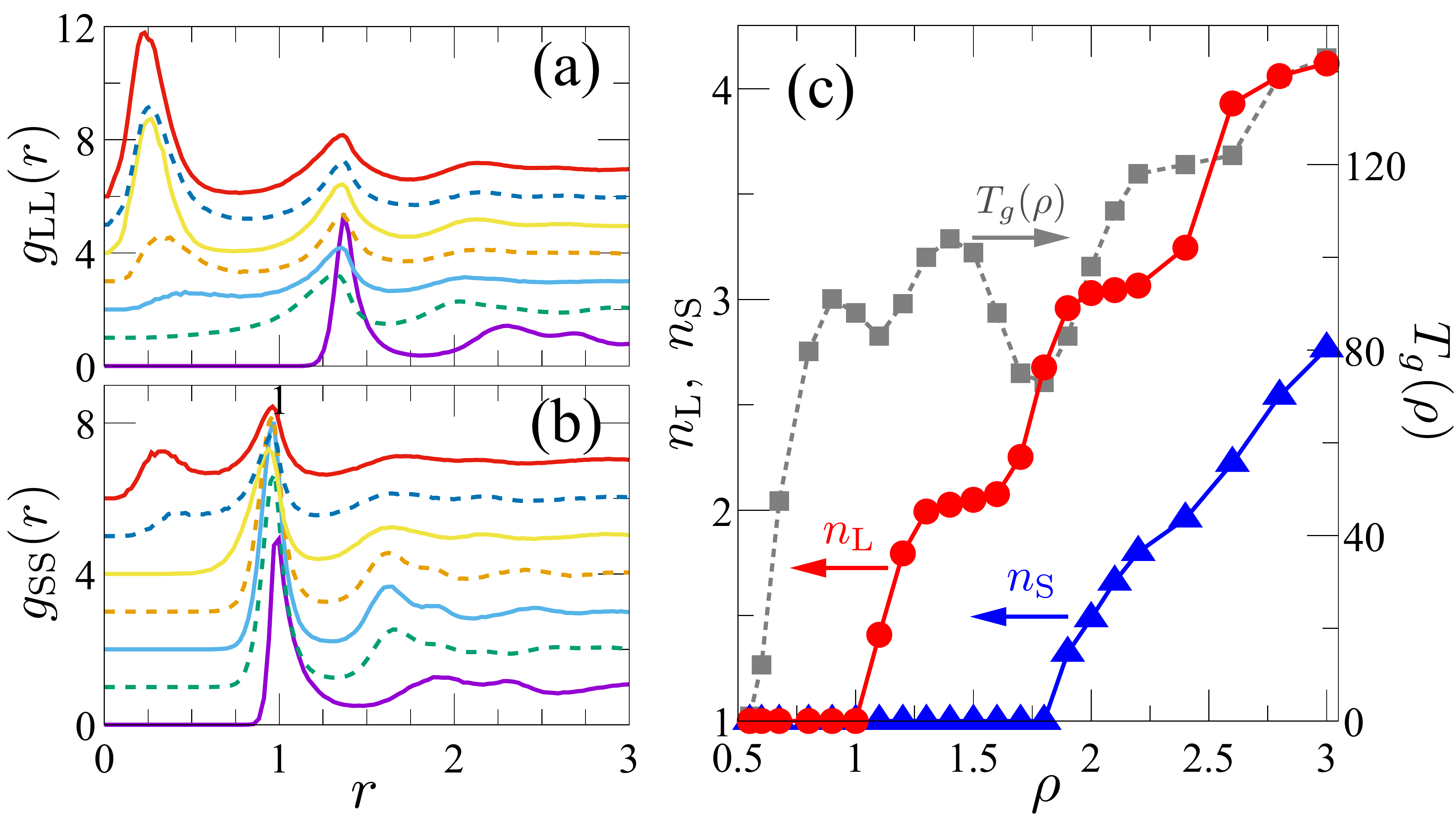}
 \end{center}
\vspace*{-0.5cm}
 \caption{
 (a)--(c) Radial distribution functions for $\alpha=3/2$ along the glass line;
 (a) $g_{LL}(r)$ and (b) $g_{SS}(r)$ at $T_{g}(\rho)$
 for $\rho = 0.675$, $1.0$, $1.1$, $1.2$, $1.4$, $1.9$, and $2.2$ from bottom to top.
 The curves are shifted vertically by 1 for clarity. 
 (c) The average number of particles per cluster of large $n_{L}$ (circles) and small
 components $n_{S}$ (triangles) along the glass line $T_g(\rho)$ (squares) for $\alpha=3/2$. 
}
 \label{Fig3}
\end{figure}
We investigate the static properties of the observed cluster glass. 
Hereafter we focus on the system with $\alpha=3/2$.
In Figs.~\ref{Fig3}(a)--3(c), the partial radial distribution functions for large
[$g_{LL}(r)$] and small particles [$g_{SS}(r)$]  at $T_g(\rho)$ for several densities are shown. 
At low density, $\rho = 0.675$, both functions show typical behaviors 
of monatomic fluids characterized by sharp nearest neighbor peaks at 
$r\approx\sigma_{a}$
($a=L$, $S$).
As the density increases, the anomaly first appears in $g_{LL}(r)$ [Fig.~\ref{Fig3}(a)]:
the nearest neighbor peak broadens and its height lowered, while the peak position shifts slightly
to smaller $r$.
At $\rho \approx 1.1$, where the first minimum is observed in $T_g(\rho)$ (Fig.~\ref{Fig2}), 
the extra peak at $r \approx 0.4$ appears and its height continuously grows
as $\rho$ increases.
This peak signals the cluster formation and 
its position matches with the bond length of the dimers shown in
Fig.~\ref{Fig1}(b).
The fact that the cluster peak is located at a distance less than $\sigma_{L}/2$
and it is well separated from the nearest neighbor peak by a deep minimum of $g_{LL}(r)$
implies that the neighboring clusters are not connected to each other.
This should be contrasted with similar peaks 
observed for the nonclustering GHP systems with $\alpha \geq 2$, 
which is caused by the interpenetration of particles as they are squashed 
by very high pressures~\cite{Zhao2011prl}.
Likewise, the cluster peak appears for small particles in $g_{SS}(r)$ but at higher densities, 
$\rho= 1.9$ [Fig.~\ref{Fig3}(b)], close to the second minimum, $\rho \approx 1.8$, of $T_g(\rho)$.
$g_{SL}(r)$ does not exhibit any cluster peak for all densities (not shown), 
meaning that the particles cluster only among the same components. 
Similar behavior was observed in a binary mixture of the GEM~\cite{Coslovich2012jcp}. 
The shape of the cluster is, however, distinct from that of the GEM for which 
the particles completely sit on top of each other so that the peak of $g(r)$ grows steeply 
at $r=0$, whereas for the GHP fluids, 
the distance between bonded particles remains finite. 
We also calculate the average number of particles per cluster for the large ($n_L$) and small ($n_S$)  components
at $T_g(\rho)$, by counting the number of adjacent particles located at $r \lesssim 0.6\sigma_{a}$
($a=$L, S) for each particle. 
These cutoff lengths correspond to the position of the deep minima of $g_{LL}(r)$ and $g_{SS}(r)$
of Figs.~\ref{Fig3}(a) and 3(b), respectively.
Figure \ref{Fig3}(c) clearly demonstrates that the large particles undergo the clustering transition from monomers
to dimers at the first minimum of $T_g(\rho)$,    
where $n_L$ steeply increases by 1, followed by a stepwise increase 
at the second and third minima of the glass line, up to $n_L \approx 4$ at the highest density. 
On the other hand, clustering of the small particles sets in at the second minima at 
$\rho\approx 1.8$ and $n_S$ also increases one by one in unison with $n_L$, although the
increases are more gradual. 
We found that particles start clustering well above $T_g(\rho)$ before the slow dynamics sets in 
but the distribution of the particle number per cluster is broader than that at $T_g(\rho)$.  
The observed semiquantized increase of $n_{L,S}$ is reminiscent of the first order
transition observed for the cluster crystals of the GEM~\cite{Zhang2010c}.
This fact may suggest the existence of a {\it liquid-liquid phase transition}
between cluster phases with different particle numbers at lower temperatures.

\begin{figure}[b]
 \begin{center}
  \includegraphics[width=\columnwidth,angle=0]{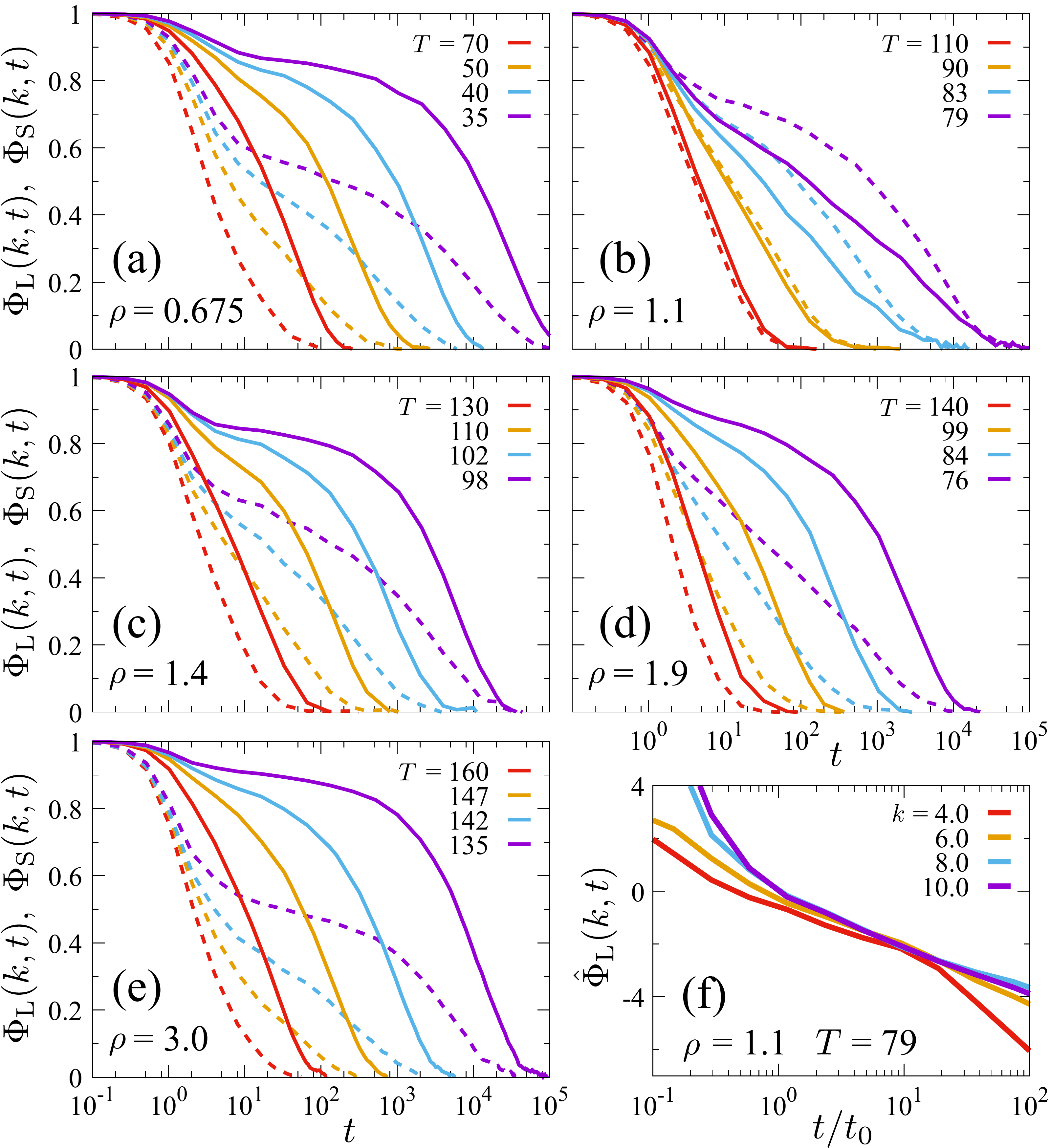}
 \end{center}
\vspace*{-0.5cm}
 \caption{
(a)--(d) 
The collective density correlation functions  for both large (solid lines) and small particles
 (dashed lines) at various temperatures (see legend).
 The wave vector $k$ is $6.0$ for large particles and $8.0$ for small particles.
 (a) $\rho =0.675$, (b) $\rho = 1.1$, (c) $\rho=1.4$, (d) $\rho=1.9$, and (e) $\rho=3.0$.
 (f) $\hat{\Phi}_{{\mbox\scriptsize L}}(k,t) = [\Fl(k,t) - f_{k}]/H_{k}^{(1)}$ at $T = 79$ for $\rho = 1.1$ with Eq.~(\ref{Eq2}) 
 for selected wave vectors around the value for which $H_{k}^{(2)} \approx 0$.
 }
 \label{Fig4}
\end{figure}
Next we carefully inspect how the slow dynamics near the glass line is affected by the cluster
formation.
In Fig.~\ref{Fig4}, the collective density correlation functions are plotted for selected
densities and several temperatures.  
Well below the clustering density, the correlation functions for both the large and small
components show the two-step relaxation characterized by a well-developed plateau 
followed by the stretched exponential relaxation [Fig.~\ref{Fig4}(a)]. 
The self part of the correlation functions $\Fselfs(k,t)$ and $\Fselfl(k,t)$ show qualitatively
similar behavior with their collective counterparts (not shown). 
We also checked that the time-temperature superposition holds, 
as expected from the mode-coupling theory (MCT) and observed for standard glass formers~\cite{Gotze2009}.
Surprisingly,  at $\rho \approx 1.1$ where the large particles start clustering, 
the plateau of $\Fl(k,t)$ disappears and the relaxation becomes logarithmic,  
whereas $\Fs(k,t)$ remains two-step shaped, as shown in Fig.~\ref{Fig4}(b).  
This peculiar relaxation is observed in the range of $1\lesssim \rho \lesssim 1.2$.
The self parts of the correlation functions, $\Fselfl(k,t)$ and $\Fselfs(k,t)$, show qualitatively
the same behavior as their respective collective counterparts.
The logarithmic relaxation has been observed both in simulations and experiments in many glassy systems, including 
the short-ranged attractive
colloids~\cite{Mallamace2000prl,*Chen2003,Pham2002,puertas2002,Zaccarelli2002b,*Sciortino2003b,Cang2003prl},
star polymers~\cite{Mayer2009,Kandar2012sm}, square-shoulder potential
fluids~\cite{Sperl2010,Das2013jcp,*Gnan2014prl}, binary mixtures with disparate particle size
ratio~\cite{Moreno2006h,*Moreno2006g}, and even in proteins~\cite{Lagi2009prl,*Chu2010sm}.
It was originally predicted theoretically by MCT as the signature
of the so-called higher order ($A_3$ or $A_4$) dynamical singularity and it is known to take place when the two or more glass
transitions compete~\cite{dawson2000,*Gotze2002}. 
We verify that the observed behavior of $\Fl(k,t)$ is genuinely due to the higher order singularity 
by fitting the data with the MCT asymptotic function 
\begin{equation}
\Fl(k, t) \sim f_k -H_{k}^{(1)}\ln (t/t_{0}) + H_{k}^{(2)}\ln^2 (t/t_{0}),
\label{Eq2}
\end{equation}
where $f_k$, $H_{k}^{(1)}$, and $H_{k}^{(2)}$ are the critical nonergodicity parameter,
and the critical amplitudes of the first and second orders, respectively~\cite{dawson2000,*Gotze2002,Gotze2009}.
In Fig.~\ref{Fig4}(f), we plot the rescaled density correlation function, 
$\hat{\Phi}_{{\mbox\scriptsize L}}(k,t) = [\Fl(k,t) - f_{k}]/H_{k}^{(1)}$.
The data demonstrate a concave-to-convex crossover across the wave vector at which $H_{k}^{(2)}=0$, where $\Fl(k,t)$ is purely
logarithmic~\cite{SM}. 
Besides, the $k$ dependence of $H_{k}^{(1)}$ is self-similar for different $T$'s and its
amplitude decreases moderately as $T$ is lowered. 
The amplitude of $H_{k}^{(2)}$ is considerably smaller than that of $H_{k}^{(1)}$~\cite{SM}.
These features are consistent with the MCT prediction 
and comparable to those of better understood model fluids~\cite{Sciortino2003b,Das2013jcp,*Gnan2014prl}. 
Therefore, we conclude that there exists a higher order singular point in the vicinity of 
the first minimum of the glass line at $\rho \approx 1.1$. 
It is natural to speculate that the two glass transition lines of monomer and cluster glasses 
intersecting at $\rho\approx 1.1$ terminate in the vicinity of but below $T_g(\rho)$ 
as is the case for the attractive and square-shoulder potential fluids~\cite{Sciortino2003b,Das2013jcp,*Gnan2014prl}. 
We emphasize that the mechanism behind the singularity in our study is not the same as 
those observed for binary mixtures with disparate size ratio
in which the smaller particles show the logarithmic relaxation~\cite{Moreno2006h,*Moreno2006g}. 
Contrarily, the singular dynamics is observed only for the large particles that undergo the cluster transition, 
while the relaxation of the smaller particles remains two-step shaped. 
Also, note that the singular dynamics is observed at the wave vectors around the nearest neighbor
peak position $k \approx 2\pi/\sigma_{L}$, implying that the length scales at play are much longer than those
for attractive glass models for which the singular dynamics is observed at much larger $k$'s~\cite{Sciortino2003b,Das2013jcp,Gnan2014prl}. 

As the density increases above $\rho \approx 1.1$, we find that $\Fl(k,t)$ gradually 
retrieves the two-step relaxation as shown in Fig.~\ref{Fig4}(c).
Interestingly, the amplitude and shape of $f_k$ remain qualitatively similar with those of the
monomer glass, in contrast with other studied systems such as the attractive colloid
glasses~\cite{Sciortino2003b,Das2013jcp,*Gnan2014prl,SM}. 
As the density reaches the second minimum of $T_g(\rho)$ at $\rho\approx 1.8$, 
where the small particles also start forming clusters,
the higher order singularity associated with the glass-glass transition is observed again, but only for
the small particles, $\Fs(k,t)$ [Fig.~\ref{Fig4}(d)]. 
$\Fl(k,t)$ stays showing two-step relaxation. 
For larger densities, $\rho > 1.8$, the two-step relaxation is recovered for $\Fs(k,t)$  [Fig.~\ref{Fig4}(e)].
The result reaffirms that the logarithmic singularity is associated with the transition from
the monomer to cluster glass phases.
At the third minimum at $\rho \approx 2.5$, where the cluster size increases further,
both $\Fl(k,t)$ and $\Fs(k,t)$ do not display logarithmic relaxation clearly.
This observation hints that the higher order singularities exist only in the vicinities of the
boundaries of monomer and dimer glass phases. 

So far, we have discussed about dynamics of only the collective part of the density correlation functions.
It is known that dynamical behavior of the self part  is qualitatively the same as
the collective counterparts for many standard glass models~\cite{Gotze2009}. 
This is also the case for the GHP fluids at the low density regime 
below $\rho\approx 1.8$, 
where the second cluster-glass transition takes place. 
The relaxation curves of both $\Fselfl (k,t)$ and $\Fselfs (k,t)$ are similar to 
$\Fl (k,t)$ and $\Fs (k,t)$, respectively, as discussed above. 
At $\rho \gtrsim 1.8$, however, dynamics of the self part becomes more complicated and decoupled
from the collective parts. 
First, not only $\Fselfs (k,t)$ but $\Fselfl (k,t)$ also demonstrates the logarithmic relaxation
when only the small particles undergo the monomer-cluster glass transition [see Fig. S4 (a) in the Supplemental Material~\cite{SM}].  
Second, the relaxation of $\Fselfs (k,t)$ is systematically facilitated and 
decouples from the collective parts as the density increases, whereas
$\Fselfl (k,t)$ retrieves the two-step relaxation [see Fig. S4 (b) in the Supplemental Material~\cite{SM}].  
The latter is reminiscent of the dynamical anomaly
observed in the binary mixture of the particles with disparate size ratio~\cite{Moreno2006h,Moreno2006g}.
In this density region, we observed a trace of particles hopping from a cluster to adjacent
clusters. 
We speculate that the observed decoupling of dynamics of the large and small particles
as well as dynamics of the self- and collective correlation functions 
is attributed to both this rare hopping event and the size disparities of the clusters. 

In summary, we reported a new type of the multiple glass transitions of the cluster forming
fluids of the generalized Hertzian potential (GHP).
The transition from one glass phase to another is signaled by deep minima of $T_g(\rho)$ and
concomitant semiquantized increase of the cluster size. 
This suggests the possible existence of the liquid-liquid phase transition between different cluster fluid phases at
lower temperatures.  
The most striking finding is the higher order dynamical singularity at the
glass-glass phase boundary. This is characterized by the logarithmic relaxation of the correlation function
which has been originally predicted by the mode-coupling theory and observed in many
glass models~\cite{Sciortino2003b,Das2013jcp,Gnan2014prl}.  
So far the liquid-liquid phase transitions have been observed mostly in
the model fluids with the anisotropy or the two length scales built in their potentials~\cite{Smallenburg2014natp,Jagla1999}.
The GHP or the ultra-soft potential fluids in general, whose potentials are isotropic and purely
repulsive, are the ideally simple and novel systems which not only help our understanding of 
the glass transition problem at a fandamental level but also may shed a new light on the study of liquid polyamorphism.

\begin{acknowledgments}
We thank D. Coslovich, A. Ikeda, H. Ikeda, and M. Ozawa for helpful discussions.
We acknowledge KAKENHI Grants
No. 24340098, 
No. 25103005, 
No. 25000002, 
and the JSPS Core-to-Core Program.
\end{acknowledgments}

%





\pagebreak

\setcounter{equation}{0}
\setcounter{figure}{0}
\setcounter{table}{0}
\setcounter{page}{1}

\renewcommand{\theequation}{{S}-\arabic{equation}}  
\renewcommand{\thefigure}{{S}\arabic{figure}}  

 \newcommand{\Hk}[1]{H_{k}^{(#1)}}
 \newcommand{\vk}{\tilde{v}(k)}

\begin{center}
\textbf{\large Supplemental Material}
\\[3pt]
\textbf{\large for}
\\[3pt]
\textbf{\large ``Cluster Glass Transition of }
\\[3pt]
\textbf{\large Ultrasoft-Potential Fluids}
\\[3pt]
\textbf{\large at High Density''}
\\[12pt]
$\text{Ryoji Miyazaki,}^{1} \ \text{Takeshi Kawasaki,}^{1}$
\\[3pt]
\text{and}
\\[3pt]
$\text{Kunimasa Miyazaki}^{1}$
\\[3pt]
\textit{\small $^{1}\text{Department of Physics, Nagoya University, }$}
\\[3pt]
\textit{\small Nagoya 464-8602, Japan}
\end{center}

\section{Fourier transform of the generalized Hertzian potential}

\begin{figure}[b]
 \begin{center}
\includegraphics[width=0.9\columnwidth,angle=0]{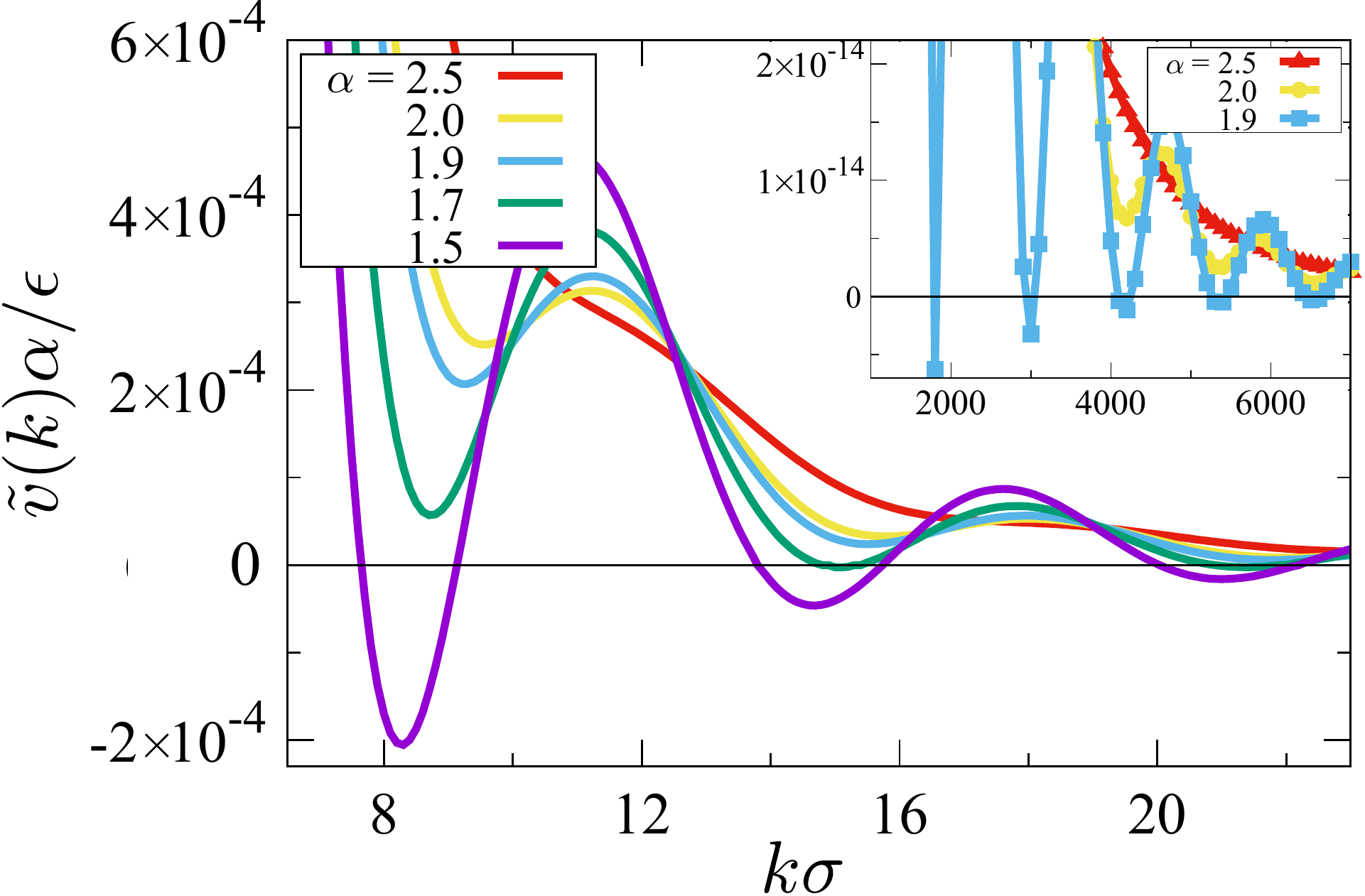}
 \end{center}
 \caption{The Fourier transform $\tilde{v}(k)$ of the generalized Hertzian potential for several $\alpha$'s.
 In the inset, the closeups of $\vk$ for $\alpha = 1.9$, 2.0, and 2.5 are shown.
}
 \label{fig:vk}
\end{figure}
We show the Fourier transformation, $\tilde{v}(k)$, of the generalized Hertzian potential
given by
\begin{equation}
v(r) = \left\{ \begin{array}{cc}
	\cfrac{\epsilon}{\alpha}\left(1-\cfrac{r}{\sigma}\right)^{\alpha} 	 &  (r < \sigma) 
	 \\
		0   &  (r > \sigma).
	       \end{array}
\right.
\end{equation}
In Fig.~\ref{fig:vk}, the results for for $\alpha = 1.5, \ 1.7, \ 1.9, \ 2.0$, and $2.5$ are plotted. 
We find that $\tilde{v}(k)$ is positive definite for $\alpha \geq 2$, but for $\alpha < 2$, 
$\tilde{v}(k)$ becomes negative for finite $k$'s. 
Exceptionally at $\alpha=2$, $\vk$  can be written analytically as
 \begin{equation}
  \vk \propto 4k\sigma - 6 \sin(k\sigma) + 2k\sigma \cos(k\sigma),
 \end{equation}
which is always positive.
Figure~\ref{fig:vk} shows that $\vk$ for the generalized Herzian potential is an oscillating
function even for $\alpha \geq 2$. 
For $\alpha < 2$,  the smallest wavevector at which $\vk$ becomes negative shifts 
to a smaller value and the amplitude of the negative dips increases as $\alpha$ decreases.

\section{Fits of the density correlation functions by the logarithmic law}

\begin{figure}[b]
 \begin{center}
  \includegraphics[width=0.9\columnwidth,angle=0]{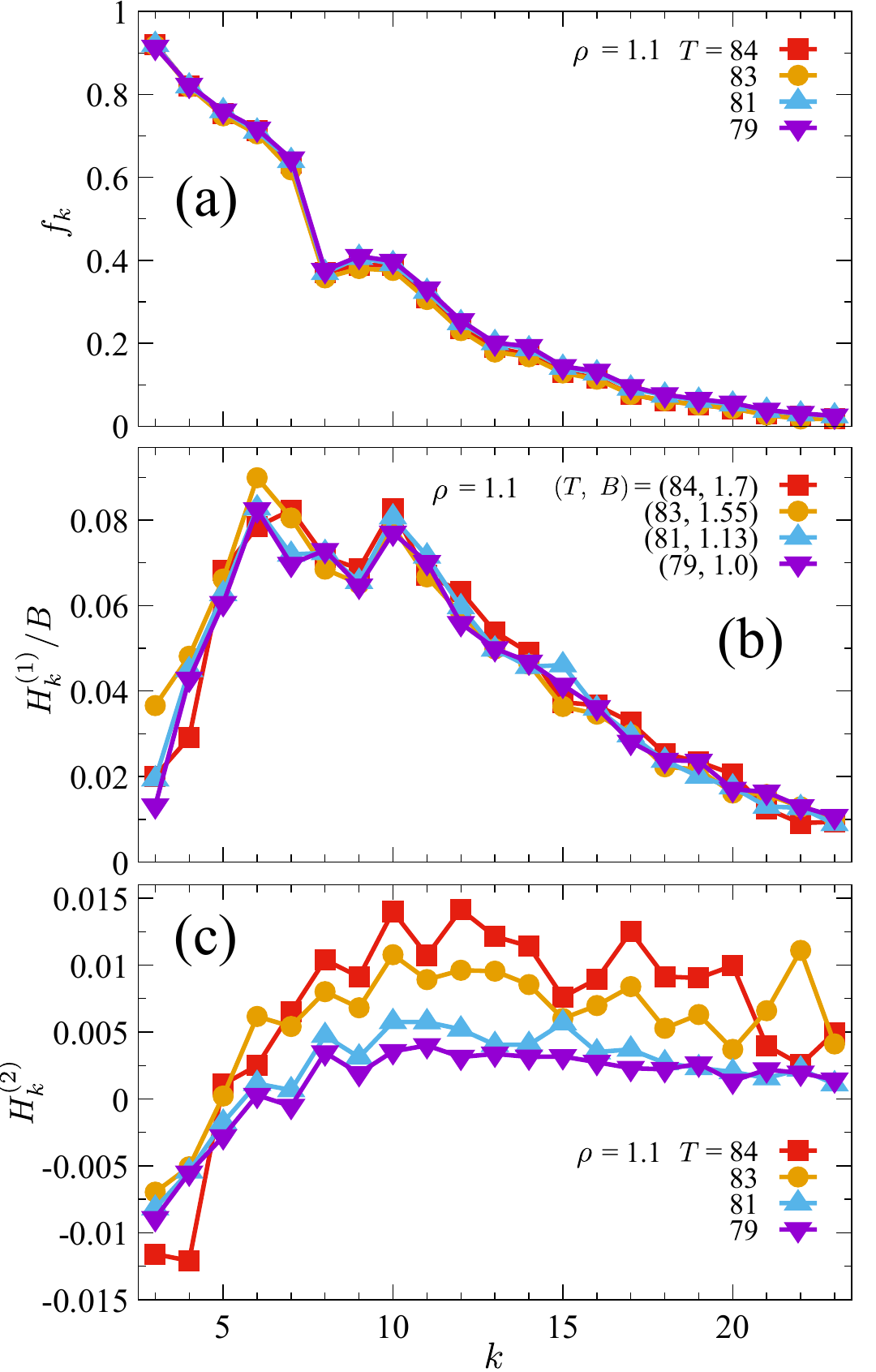}
 \end{center}
 \caption{The values of (a) $f_{k}$, (b) $\Hk{1}/B$, and (c) $\Hk{2}$ for the systems at $\rho = 1.1$ and selected low temperatures (see legend), obtained from fits of $\Fl(k,t)$ to the asymptotic logarithmic law, Eq.~(\ref{eq:hos_FA}).
 The value of $t_{0}$ is $4$, $5$, $6$, and $7$ for $T = 84$, $83$, $81$, and $79$, respectively.
}
 \label{fig:hos}
\end{figure}

\begin{figure}
 \begin{center}
  \includegraphics[width=0.9\columnwidth,angle=0]{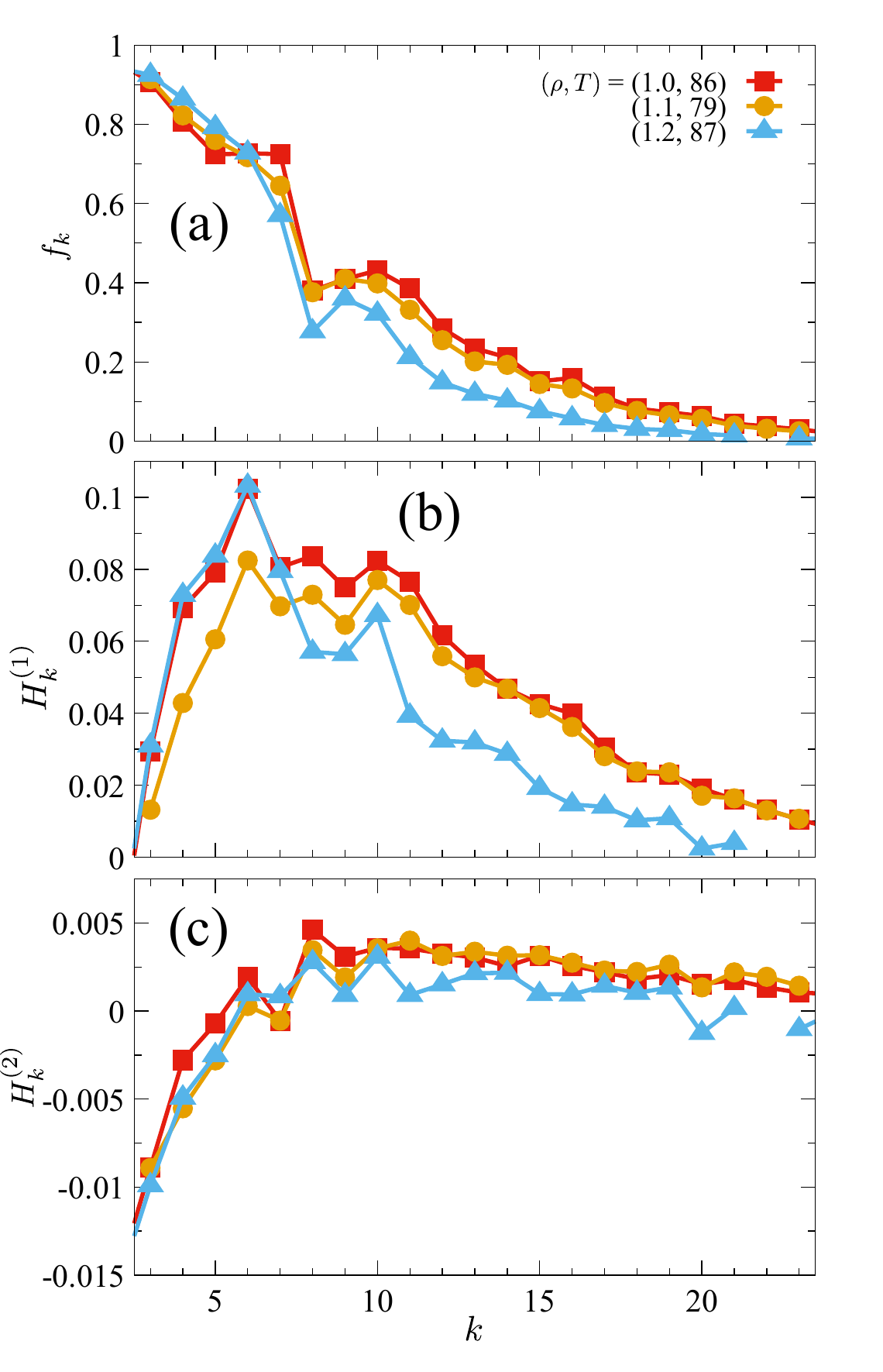}
 \end{center}
\caption{
The values of (a) $f_{k}$, (b) $\Hk{1}$, and (c) $\Hk{2}$ at $(\rho, T) = (1.0, 86)$, $(1.1, 79)$, and $(1.2, 87)$, 
obtained from fits of $\Fl(k,t)$ to the asymptotic logarithmic law, Eq.~(\ref{eq:hos_FA}).
 }
 \label{fig:hos_2}
\end{figure}
The collective density correlation function for the large particles, $\Fl(k,t)$, at the density
$\rho \simeq 1.1$ exhibits the logarithmic decay as shown in Fig.~2(b) in the main text.
Following the previous works~\cite{sm_Sciortino2003b,sm_Das2013jcp,*sm_Gnan2014prl}, we fit the data with the
polynomials of $\ln t$ given by 
\begin{equation}
\Fl(k,t) \sim f_{k} - \Hk{1} \ln (t/t_0) + \Hk{2} \ln^{2} (t/t_0),
 \label{eq:hos_FA}
\end{equation}
where $t_0$, $f_k$, $\Hk{1}$, and $\Hk{2}$ are the fit parameters. 
These fit parameters obtained for $\rho = 1.1$ are plotted as functions of $k$ in Fig.~\ref{fig:hos}(a)--(c). 
We determined $t_0$ in such a way that $f_{k=6.0} \approx 0.7$.
$f_{k}$'s for different temperatures collapse on a single function as shown in Fig.~\ref{fig:hos}(a).
In Figure~\ref{fig:hos}(b), we plot $\Hk{1}$ for several temperatures scaled with a $k$-independent
constant $B$. This result demonstrates that $\Hk{1}$ is self-similar in $k$ and can be decomposed 
as $\Hk{1} = h_{k}B(T)$.  
This is consistent with the prediction of MCT which claims that $B(T)$ is independent of $k$~\cite{sm_dawson2000,*sm_Gotze2002,sm_Gotze2009}.
We find that $B(T)$ is a moderately decreasing function of the temperature.
The parameter $\Hk{2}$ is shown in Fig.~\ref{fig:hos}(c).
We find that $\Hk{2}$ is negative for small $k$ and positive for large $k$.
$\Hk{2}$ vanishes at $k \simeq 6$ which is close to the first peak of the static
structure factor for the large particles. 
Therefore, if $\hat{\Phi}_{L}(k,t) \equiv \left[ \Fl(k, t)- f_k\right]/\Hk{1}$ is plotted as a function of $\ln t$, 
the function shows the concave-to-convex transition as $k$ increases.
Furthermore, the amplitude of $\Hk{2}$ is much smaller than $\Hk{1}$.
This observation is consistent with MCT according to which
$B(t)$ (and thus $\Hk{1}$) is proportional to $\sqrt{\varepsilon} = \sqrt{|T/T_c-1|}$, where $T_c$ is the MCT transition
temperature, and $\Hk{2}$ is the order of $\varepsilon$~\cite{sm_dawson2000,*sm_Gotze2002,sm_Gotze2009}.
These results are also comparable to the results of the previous studies for the short-range
attractive and square-shoulder potential fluids, except for the fact that 
the logarithmic decay is observed at much shorter wavevectors as we discussed in the main
text~\cite{sm_Sperl2003,sm_dawson2000, sm_Zaccarelli2002b,*sm_Sciortino2003b, sm_Moreno2006h, *sm_Moreno2006g, sm_Mayer2009, sm_Das2013jcp, *sm_Gnan2014prl}. 

In order to see how the glass properties change as one crosses this dynamical singular point, 
we evaluate the fit parameters slightly above and below $\rho =1.1$, following the same fitting procedure. 
Figure~\ref{fig:hos_2} demonstrates the $k$-dependence of $f_k$, $\Hk{1}$, and $\Hk{2}$ at $\rho = 1.0$, $1.1$, and $1.2$. 
No qualitative difference between the curves for different densities is found, 
although the results for $\rho = 1.2$ are slightly smaller than the others.
This result is in stark contrast with the case of the short-range attractive colloids
for which $f_k$ discontinuously changes in the vicinicty of the singular point~\cite{sm_puertas2002,sm_Zaccarelli2002b}.
Note, however, that the latter is not a generic feature of the dynamical singularity but it is
due to the formation of strong bonds of the colloidal particles caused by the short-ranged and strong attractive potential.
Acording to the original formulation of the mode-coupling theory, the higher order dynamical
singularity is not necessarily acomapnied by the discontinous changes in the critical parameters~\cite{sm_Gotze2009}. 

\section{Self density correlation functions in the high density region}

\begin{figure}[htb]
 \begin{center}
\includegraphics[width=1.0\columnwidth,angle=0]{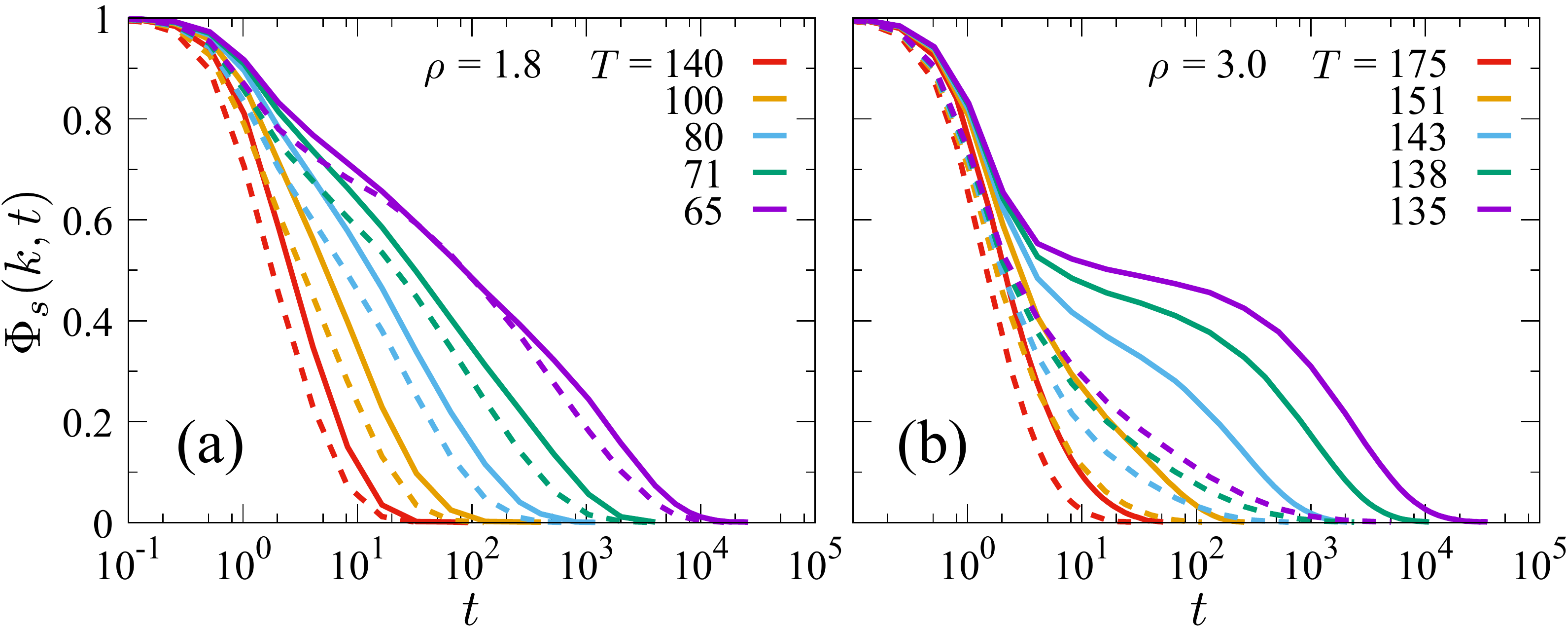}
 \end{center}
 \caption{
 The self density correlation functions for (a) $\rho = 1.8$ and (b) $3.0$ at several temperatures.
 The solid and dashed lines are for large (L) and small (S) particles, respectively. 
 The wavevectors $k$ are $6.0$ for large particles and $8.0$ for small particles.
}
 \label{fig:Fs}
\end{figure}
Figure~\ref{fig:Fs}  shows the self density correlation functions for the large and small particles, 
$\Fselfl(k,t)$ and $\Fselfs(k,t)$, for $\rho = 1.8$ and $3.0$ for several temperatures.
The wavevectors have been chosen to be $k=6.0$ and $8.0$ for the large and small particles, respectively.

%

\end{document}